\documentclass[%
 reprint,
superscriptaddress,
 amsmath,amssymb,
 aps,
prl,
]{revtex4-2}
\renewcommand{\selectlanguage}[1]{} 
\usepackage{graphicx}
\usepackage{dcolumn}
\usepackage{bm}
\usepackage{hyperref}
\usepackage{textcomp}
\usepackage{amsmath}
\usepackage{mathtools}
\usepackage{caption}

\begin{document}

\preprint{APS/123-QED}

\title{Dimensionality reduction of optically generated vortex strings in a charge density wave }

\author{Sajal Dahal}
\affiliation{%
Department of Physics, Arizona State University, Tempe, AZ, USA
}%
  
\author{Alex H. Miller}%
\affiliation{%
Department of Physics, Arizona State University, Tempe, AZ, USA 
}%

\author{Viktor Krapivin}%
\affiliation{%
PULSE Institute, SLAC National Accelerator Laboratory, Menlo Park, CA, USA 
}%
\affiliation{%
Department of Physics, Stanford University, Stanford, CA, USA 
}%
\affiliation{Stanford Institute for Materials and Energy Sciences, SLAC National Accelerator Laboratory, Menlo Park, CA 94025, USA}

\author{Gal Orenstein}%
\affiliation{%
PULSE Institute, SLAC National Accelerator Laboratory, Menlo Park, CA, USA 
}%
\affiliation{Stanford Institute for Materials and Energy Sciences, SLAC National Accelerator Laboratory, Menlo Park, CA 94025, USA}

\author{Ryan A. Duncan}%
\affiliation{%
PULSE Institute, SLAC National Accelerator Laboratory, Menlo Park, CA, USA 
}%

\author{Nicholas Leonard}%
\affiliation{%
Department of Physics, Arizona State University, Tempe, AZ, USA 
}%
  
\author{Matthew J. Hurley}%
\affiliation{%
Department of Physics, Arizona State University, Tempe, AZ, USA
}%

\author{Jade Stanton}%
\affiliation{%
Department of Physics, Arizona State University, Tempe, AZ, USA
}%
\affiliation{Present Address: Department of Applied Physics, Stanford University, Stanford, CA, USA}

\author{Roman Mankowsky}%
\affiliation{%
SwissFEL, Paul Scherrer Institute, Villigen, Switzerland 
}%

\author{Henrik Lemke}%
\affiliation{%
SwissFEL, Paul Scherrer Institute, Villigen, Switzerland 
}%

\author{Anisha Singh}%
\affiliation{%
Department of Applied Physics, Stanford University, Stanford, CA, USA 
}%
\affiliation{Stanford Institute for Materials and Energy Sciences, SLAC National Accelerator Laboratory, Menlo Park, CA 94025, USA}

\author{Ian Fisher}%
\affiliation{%
Department of Applied Physics, Stanford University, Stanford, CA, USA 
}%
\affiliation{Stanford Institute for Materials and Energy Sciences, SLAC National Accelerator Laboratory, Menlo Park, CA 94025, USA}

\author{Mariano Trigo}%
\affiliation{%
PULSE Institute, SLAC National Accelerator Laboratory, Menlo Park, CA, USA 
}%
\affiliation{Stanford Institute for Materials and Energy Sciences, SLAC National Accelerator Laboratory, Menlo Park, CA 94025, USA}

\author{Samuel W. Teitelbaum}%
\email{samuelt@asu.edu}
\affiliation{%
Department of Physics, Arizona State University, Tempe, AZ, USA
}%




\date{\today}

\begin{abstract}
In phase transitions, mesoscale structures such as topological defects, vortex strings, and domain walls control the path towards equilibrium, and  thus the functional properties of many active devices. In photoinduced phase transitions driven by femtosecond laser excitation, the temporal (pulse duration) and spatial (penetration depth) structure of the optical excitation present opportunities for control and creating structures with unique topologies. By performing time-resolved optical pump, x-ray probe experiments on the CDW system Pd-intercalated ErTe$_{3}$, we gain access to the nanoscale dynamics of the mesoscale topological features (vortex strings) produced after a quench, which have a different apparent dimensionality than the topological defects predicted from the bulk system. We show that these vortex strings persist for much longer than the electronic recovery time. The critical exponent obtained from power-law scaling of the intensity as a function of wavevector shows a reduction in the effective dimensionality of the topological defects in the system, corroborated by time-dependent Ginzburg-Landau simulations. Our results demonstrate a novel pathway to use light to control the dimensionality and orientation of topological defects in quantum materials, which could be used to stabilize competing quantum states.

\end{abstract}

\maketitle
 

The functional properties of materials are often determined by their approach towards equilibrium from a highly nonequilibrium state, from the formation of defects in metals to strongly driven quantum materials \cite{orenstein_dynamical_2025,basov_towards_2017,hiroshi_6_2014}. In light-induced phase transitions in condensed matter systems, it is an open question as to when a system's evolution is predominantly coherent (e.g. occurs via the coherent, uniform motion of atoms) or incoherent (e.g. driven by thermal motion) \cite{wall_ultrafast_2018,johnson_all-optical_2024,han_exploration_2015}. An incoherent evolution into an ordered state must produce topological defects at boundaries between domains, as described by e.g. the Kibble-Zurek mechanism \cite{kibble_topology_1976,zurek_cosmological_1985}, and also predicted by renormalization group theory \cite{lal_universal_2020,mondello_scaling_1992, mitrano_ultrafast_2019, vogelgesang_phase_2018}. 

When the timescale of a quench (e.g.~the effective temperature change) approaches the fundamental motion timescale of the order parameter (for example a phonon period in a structural phase transition) scaling models break down, because there is a fundamental ``speed limit" in the system. Optically driven phase transitions also have a fundamental length scale defined by the optical penetration depth \cite{trigo_ultrafast_2021}. The role of the temporal structure of light in controlling phase transitions has been explored in detail \cite{hsieh_evidence_2014,huber_coherent_2014,trigo_ultrafast_2021,de_la_pena_munoz_ultrafast_2023}, but the \emph{spatial} structure of light is often assumed to play a less important role, due to the long wavelength of light relative to the size of the mesoscale features in a typical quantum system \cite{zong_evidence_2019,wandel_enhanced_2022}. In this paper, we show experimentally that contrary to this assumption, the spatial structure of light has important consequences for how the quench proceeds \cite{kong_fate_2025}.

Experimental access to these length (nm) and time (ps) scales is critical to delineate between different mechanisms and pathways of relaxation and track their time-dependent structure. By combining time-resolved x-ray scattering using an x-ray free electron laser and time-dependent Ginzburg-Landau simulations, we show that the effective dimensionality of the photoinduced topological defects of the CDW order in a rare-earth tritellruide is reduced. Furthermore, these defects persist for a timescale that is orders of magnitude longer than the electronic quench dynamics.

\begin{figure}[ht!]
    \centering
    \includegraphics[width=1\linewidth]{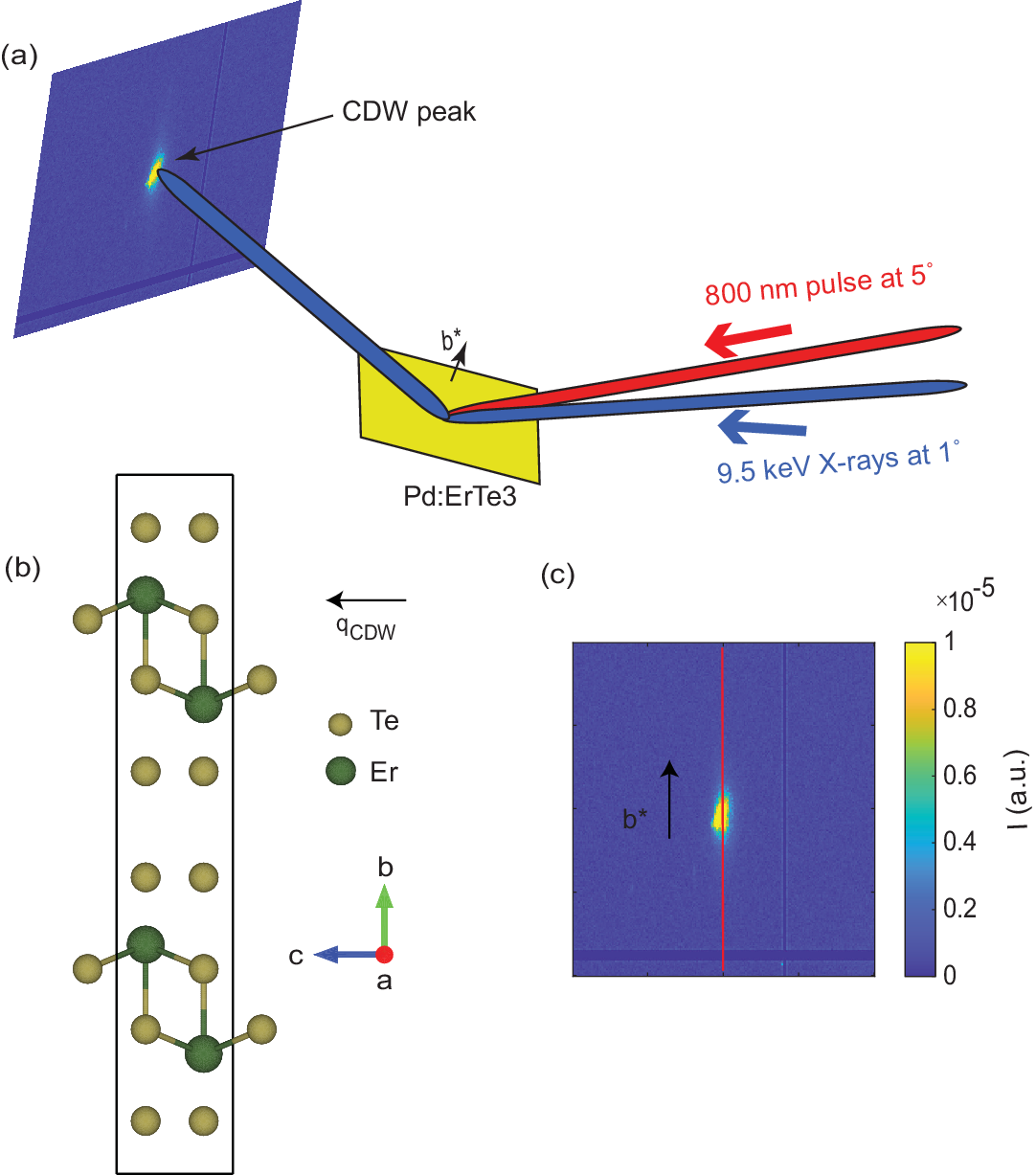}
    \caption{ (a) A schematic of the pump-probe experimental setup. An optical laser ($\lambda = 800$ nm) excites a single crystal of Pd:ErTe$_3$ and a delayed 9.5 keV x-ray probe pulse diffracts from it. (b) Crystal structure of ErTe$_{3}$ viewed in the bc plane, with the approximate $q_{\mathrm{CDW}}$ direction, and the surface normal in the vertical direction. (c) Detector image for the $(1 \ 3 \ 1 -\mathrm{q_{\mathrm{CDW}}})$ CDW peak. The red line in fig. \ref{fig:data}(b) is where the linecut is taken, and is approximately parallel to the surface normal (b$^*$).}
    \label{fig:setup}
\end{figure}

The system used in this study is Pd-intercalated ErTe$_3$. ErTe$_{3}$ is a member of the rare-earth tritelluride (RTe$_{3}$) family, a series of prototypical charge density wave (CDW) forming materials \cite{orenstein_dynamical_2025, ru_charge_2008}.  The tritellurides have an orthorhombic (nearly tetragonal) crystal structure where the CDW forms with wavevector along the $c$-axis, where the $b$-axis is the van der Waals direction (fig.~\ref{fig:setup}(b)). Substitution of the rare-earth element R tunes the CDW transition temperature, with heavier rare-earths having a lower transition temperature. In addition to the $c$-axis CDW, for heavier R ions, RTe3 undergoes a second transition at lower temperatures with a perpendicular CDW state with wavevector along the $a$-axis.  Pd-intercalation is also known to increase interlayer screening, further reducing the $c$-axis CDW transition temperature \cite{iguchi_anomalous_2024}. The tunability, controllable presence of other competing phases, and relatively high critical temperatures makes the $c$-axis CDW in the $\mathrm{RTe_3}$ system a compelling platform for studying phase transitions in quantum materials.


We performed time-resolved optical pump, x-ray diffraction probe on Pd-intercalated ErTe$_{3}$ using the Bernina endstation at SwissFEL \cite{ingold_experimental_2019, prat_compact_2020}. Single-crystal samples were grown by the self-flux method \cite{singh_emergent_2024}, and have a Pd-intercalation level of 1\%. The crystal was cleaved just before the experiment under a continuous dry nitrogen stream to prevent oxidation. A schematic of the experimental setup is shown in fig. \ref{fig:setup} (a). The normal surface of the sample is along the [0 1 0] ($b^*$) direction. The crystal was irradiated with a p-polarized, 35 fs FWHM, 800 nm wavelength pump incident at $5^\circ$ with respect to the incoming x-ray beam.  The optical laser on the crystal has an incident fluence of $2\  \mathrm{mJ/cm^2}$ focused onto a 0.3 mm $\times$ 0.4 mm FWHM spot projected along the surface. We probed the dynamics with 50 fs FWHM, 9.5 keV x-ray pulses at 1$^\circ$ grazing angle relative to the surface plane, producing a $1/e$ x-ray penetration depth of $\delta = 100 \ \mathrm{nm}$. The x-rays were focused onto a 0.57 mm $\times$ 0.2 mm spot projected onto the sample surface. The sample azimuthal angle $\phi$ was tuned such that the Bragg condition was met for the $(1 \ 3 \ 1-q_{\mathrm{CDW}})$ CDW satellite peak. The scattered x-rays were collected using a Jungfrau 1M area detector \cite{mozzanica_prototype_2014} placed 0.5 meters from the sample. Scans of the x-ray/laser time delay were made from $t = -1$ ps (before laser excitation) to $t = +5 \ \mathrm{\mu}$s (after laser excitation) with a logarithmic time spacing from 100 fs to 1 $\mu$s following laser excitation. The sample was kept below the CDW ordering temperature of $T_c = 200 \mathrm{K}$ \citep{singh_emergent_2024} at $T = 80\ \mathrm{K}$ using a nitrogen cryostream (Oxford Instruments). Experiments were also conducted at $T = 100\ \mathrm{K}$ and $T = 150\ \mathrm{K}$. These experiments did not show any qualitative differences with variation of the sample temperature, and the data presented in this work was taken at 80 K.

\begin{figure}[ht]
    \centering
    \includegraphics[width=1\linewidth]{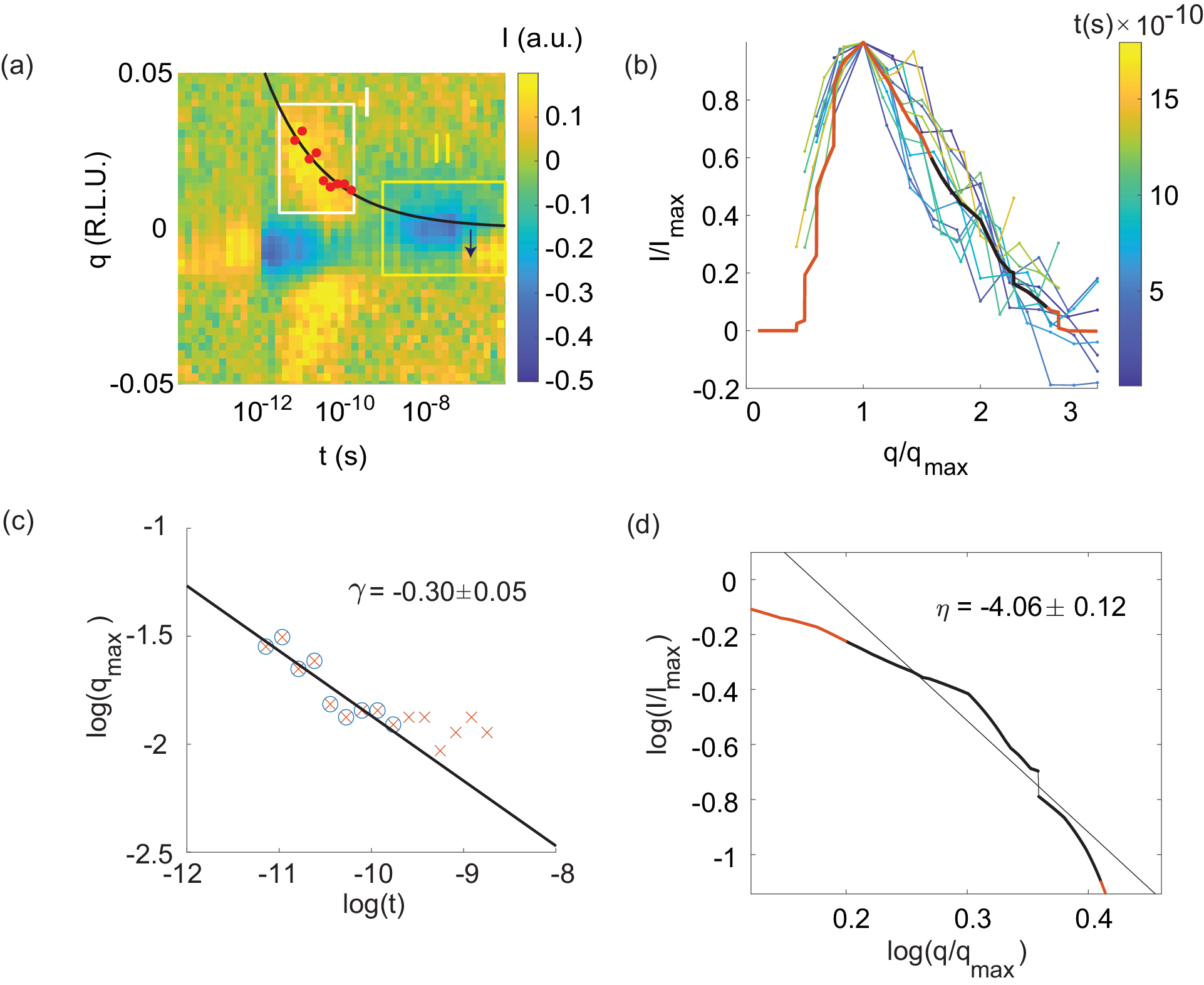}
    \caption{ (a) The linecut with the ``domain wing" (region I, white box), the wavevector shift of the intensity change near $q = q_{\mathrm{CDW}}$ (black arrow in region II, yellow box) and the $\gamma$ power-law fit (black curve). Red circles are used for the fit. The $t$-axis is spaced logarithmically. (b) A linear plot of the normalized intensity vs. normalized wavevector at different time delays with the interpolated average in orange and black, where the black portion represents the part used for the $\eta$ power-law fit. (c) Log-log plot of the power-law fit in (a) with the critical exponent showing subdiffusive behavior. Red crosses indicate all points extracted and blue circles enclose the points used for the fit (corresponding to the red points in (a)). The points not used in the fit are subsumed by the Bragg peak. (d) Log-log plot of (b), showing the interpolated average in orange and black with the black portion being used to extract the $\eta$ critical exponent (same color scheme as (b)). The discontinuity in the black region is due to a step in the averaging.}
    \label{fig:data}
\end{figure}

We take a vertical linecut along the $b^*$-direction in reciprocal space (approximately along the surface normal) passing through the CDW peak as indicated by the red line in fig.~\ref{fig:setup}(c). We then map the change in intensity along this linecut as a function of time after photoexcitation, $\Delta I/I(\bm{q},t)$ (fig. \ref{fig:data}(a))). The calibration of reciprocal space and known detector distance, pixel size, lattice constant, and x-ray energy allows conversion from detector pixel units to reciprocal lattice units (R.L.U.).

A false color plot of the differential time-dependent scattering linecut $\Delta I/I_0(\mathbf q)$ is shown in fig. \ref{fig:data} (a), showing two key features of the data. The first key feature in the data is the increase in intensity at wavevectors away from the CDW peak from $t = 3$ ps to $t = 200$ ps (region I in fig. \ref{fig:data} (a)), which starts at high wavevector and moves to lower wavevector at increasing time delays. This feature is due to the antiphase domain wall formed by strong photoexcitation \cite{trigo_ultrafast_2021,kong_fate_2025}. The second feature is the shift in the wavevector of the CDW peak along the $b^*$ axis near $q=q_{\mathrm{CDW}}$ from $t = 1$ ns to $t = 1 \ \mu$s (black arrow, region II in fig. \ref{fig:data} (a)) due to persistent low-energy phase modes, a slow linear phase shift in the penetration depth direction that results in a shift in the CDW peak in reciprocal space. We will discuss the analysis of the early-time, high-wavevector feature first. 


The intensity of x-ray scattering near the CDW peak is proportional to the structure factor $S(q,t)=|\tilde{\phi}(\mathbf{q},t)|^2$, where $\tilde{\phi}(\mathbf{q},t)$ is the Fourier transform of the complex order parameter $\phi(\mathbf{r},t)$ \cite{bray_growth_1994}. Therefore, we can interpret the movement of the scattering to lower wavevector at longer time delays as a coarsening process, where the domain structure becomes larger in real space over time. 

We can further classify this coarsening process using predictions from  Ginzburg-Landau field theory of phase transitions and renormalization group methods of predicting critical exponents. These models predict the structure factor quenched into an ordered phase after a symmetry breaking phase transition follows (1) universal scaling (the structure factor as a function of time falls onto a universal curve when scaled, normalized by intensity and wavevector) and (2) the scaling relationship is a power-law \cite{orenstein_dynamical_2025, bray_growth_1994}:
\begin{equation}\label{eq:structure-scaling}
    S(q,t) \sim L(t)^dF[qL(t)], 
\end{equation}
where $F$ is a universal function and $d$ is the dimensionality of the system. $L(t)$ is the time-dependent universal length scale of the form $(At)^{-\gamma}$, where $A^{-\gamma}$ is a proportionality constant and $\gamma$ determines the type of coarsening of the domain walls (superdiffusive if $\gamma>0.5$, diffusive if $\gamma=0.5$, or subdiffusive if $\gamma<0.5$) \cite{orenstein_dynamical_2025,bray_growth_1994,mazenko_instability_1985}. We find that $\gamma=-0.30 \pm 0.05$ (see fig~\ref{fig:data}(c)), indicating that the growth is subdiffusive, as we found previously for LaTe$_3$ \cite{orenstein_dynamical_2025}.

The scaling of the structure factor with wavevector reveals key mechanistic details about the topological defects that mediate the phase transition. The scattering intensity as a function of wavevector $q$ should follow:
\begin{equation} \label{eq:eta-scaling}
    |S(q)| = Bq^\eta
\end{equation}
when $q \gg q_{max}$, where $q_{max}$ is the wavevector of the maximum intensity $I_{max}$ for each time delay $t$. The critical exponent $\eta$ should be $\eta = -(n + d)$, where $n$ is the dimensionality of the order parameter (here $n=2$ for an incommensurate CDW) and $d$ is the dimensionality of the system ($d=3$) \cite{mondello_scaling_1990,orenstein_dynamical_2025}.

To verify that the data collapses onto a universal curve, we plot the normalized intensity $I(\bm{q})/I_{\mathrm{max}}$ as a function of the normalized wavevector $q/q_{\mathrm{max}}$, shown in fig. \ref{fig:data} (b). Within the region where the scattering is clearly visible in the linecut (region I in fig. \ref{fig:data}(a)), the data collapses. The data is expected to not scale at low $q$ at long times due to the length-scale in that region being comparable to other measurement parameters such as the laser penetration depth \cite{orenstein_dynamical_2025, trigo_ultrafast_2021}.

We extract the exponent $\eta$ in eq. \ref{eq:eta-scaling} by fitting the interpolated average of all times (orange and black curve where black is used for the fit in fig \ref{fig:data} (b)) to a power law. We extract a power-law of $\eta = -4.06 \pm 0.12$, which is close to $\eta = -4$, rather than the $\eta = -(n + d) = -5$ predicted from the dimensionality of the system, and observed previously in LaTe$_3$ \cite{orenstein_dynamical_2025,bray_universal_1993,mondello_scaling_1990,mondello_scaling_1992}. The average across multiple experimental runs was $\eta = -4.04 \pm 0.43$. Tuning temperature from 80 K to 135 K did not have any nontrivial effect on the values of the critical exponents, as expected from universal scaling dynamics. 

The main discrepancy between the experimental results observed here, renormalization group predictions and previously observed results in LaTe$_3$ is the power-law scaling of the intensity with wavevector, $\eta$. The system is three-dimensional ($d=3$) with a complex order parameter ($n=2$). We hypothesize that the system takes on reduced dimensionality ($d=2$) as a consequence of the initial photoexcitation process, specifically at length scales comparable to the optical penetration depth. Previous experiments in the photoinduced phase transition in LaTe$_3$ observed phase modes and vortex strings at much higher wavevectors (shorter length scales) than the optical penetration depth \cite{orenstein_dynamical_2025}. Here, we track the spatial texturing of the order parameter on the same length scale as the optical penetration depth, which we hypothesize can create a fundamentally new regime for the spatial texturing of the order parameter.

In order to investigate this hypothesis further, we simulate our experiment with a time-dependent Ginzburg-Landau model which explicitly includes the spatial and temporal dependence of the optical excitation pulse in three dimensions. In this model, the CDW order parameter is modeled as a complex number $\phi(\bm{r},t)$ on a three-dimensional grid as a function of time.
\begin{figure*}[t]
    \centering
    \captionsetup{width=\linewidth} 
    \includegraphics[width=\linewidth]{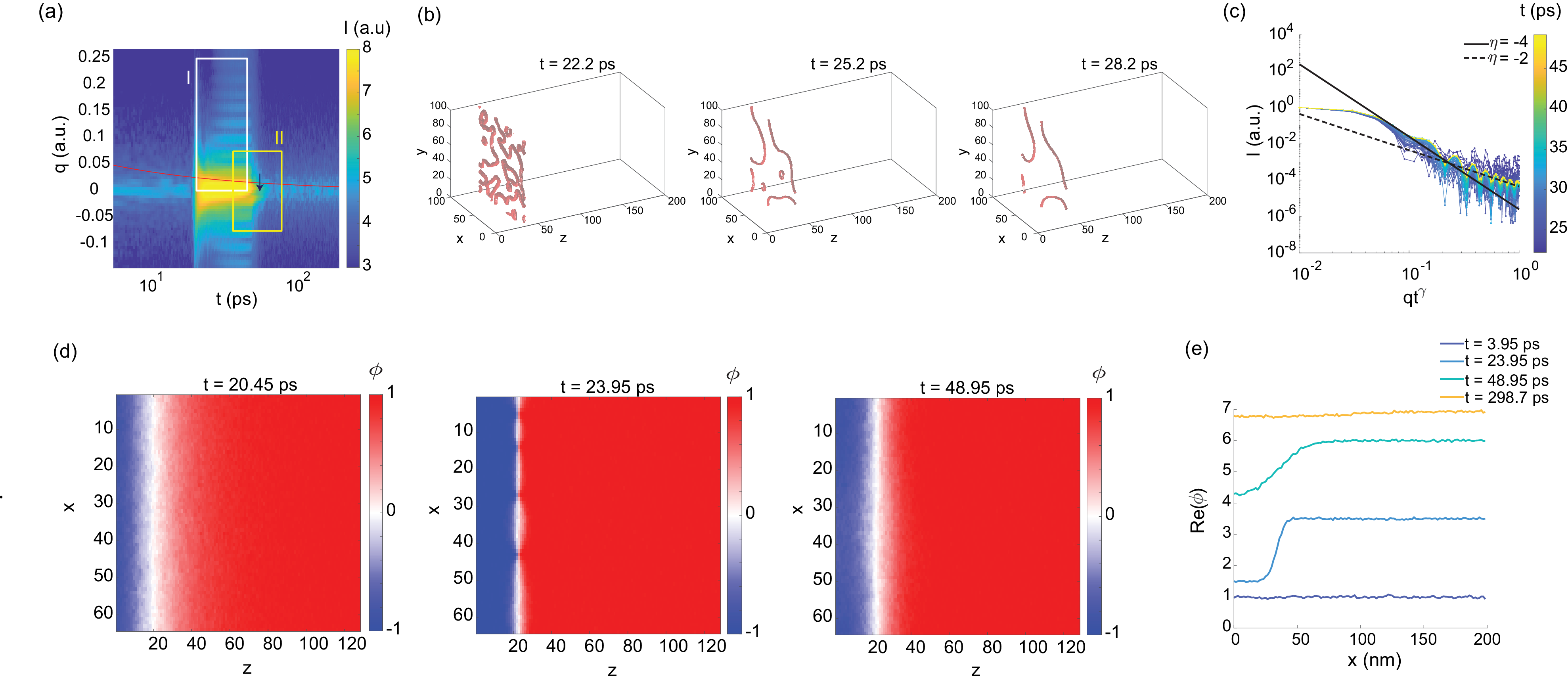}
    \caption{ (a) Linecut of the simulated diffraction with the regions I (white box) and II (yellow box including the black arrow showing wavevector shift of the intensity change near $q = q_{\mathrm{CDW}}$) corresponding to that of fig. \ref{fig:data} (a). The red curve represents the $\gamma$ power-law fit, with $\gamma=-0.5$. (b) Isosurface plot showing the formation and annihilation of vortex strings. The full movie is shown in isosurface\_movie.mp4. (c) Differential intensity vs. wavevector (from region I) times $t^\gamma$ where $\gamma = -0.5$. The solid black line shows a power-law of $\eta = -4$ for the topological defects (coming the part of region I that does not overlap with region II), and the dashed line shows $\eta = -2$ for persistent phase modes. (d) Real space plot in pixels of the order parameter showing the formation of vortices. The full movie is shown in realspace\_movie.mp4. (e) The real part of order parameter vs. depth at different time delays. The $y$-axis is offset at different times for clarity.}
    \label{fig:simulations}
\end{figure*}

We use the nonlinear Klein-Gordon equation \cite{mondello_scaling_1992, yusupov_coherent_2010} to model the evolution with a complex order parameter $\phi(\mathbf{r},t)\equiv \phi$: 
\begin{equation}
\label{eq:GL-model}
\begin{aligned}
\frac{\partial^2\phi}{\partial t^2} = \frac{\Omega_0^2}{2}\bigg(\big(1-\alpha(\bm{r},t) - |\phi|^2\big) \phi \\
+ \xi^2\nabla^2\phi\bigg) - \Gamma\frac{\partial \phi}{\partial t } + \zeta(\mathbf{r},t)
\end{aligned}
\end{equation}

where $\Gamma$ is a phenomenological damping parameter ($\simeq \frac{0.12} {\Omega_0^2}$), $\xi$ is the spatial stiffness of the order parameter (1.2 nm), and $\zeta(\bm{r},t)$ is a random Langevin force simulating thermal noise \cite{trigo_ultrafast_2021,orenstein_dynamical_2025}.

Photoexcitation modifies the equilibrium amplitude mode frequency $\Omega_0$ ($\simeq 2\pi \times 1.7$ THz), and its effect is simulated as a change in the interatomic potential, $\alpha(\bm{r},t$), which is proportional to the spatial and time-dependent photoexcited carrier energy density:
\begin{equation}
    \alpha(\bm{r},t) = \eta_0 e^{-t/\tau}\times e^{-z/\delta}
\end{equation}
where $\tau$ is the excitation lifetime (1 ps), $\delta$ is the optical penetration depth (60 nm) and $\eta_0$ is proportional to the the fluence ($1.25$). We assume the beam is uniform in the $x$ and $y$ directions, as the optical spot size is much larger than all other length scales in the problem.

We carry out the simulation in three dimensions, on a $64 \times 64 \times 128$ spatial grid and 100 nm $\times$ 100 nm $\times$ 200 nm volume from $t = -1$ ps to $t = 299$ ps with 50 fs time steps. We thermalize the system by running the simulation $t = 20$ ps before excitation. In the $z$ dimension, we impose non-periodic boundary conditions and evaluate the laplacian in equation \ref{eq:GL-model} using a discrete cosine transform (see supplemental information in \cite{trigo_ultrafast_2021}), and use periodic boundary conditions in the $x$ and $y$ dimensions \cite{orenstein_dynamical_2025}. The x-ray penetration depth is set to $\delta_{\mathrm{xray}}= 200$ nm.

Figure \ref{fig:simulations} (a) shows a false color plot of the time-dependent scattering linecut along the $b^*$-axis from the simulation. Region I (in white) shows the formation of antiphase domain walls after photoexcitation and region II (in yellow) shows the low energy phase modes which persist for much longer than the topological defects. The mismatch in timescales compared to the experiment arises from the value of the parameters $\xi$ and $\Omega_0$ in eq. \ref{eq:GL-model} which scale the relaxation time. Region II in the simulation also does not show negative intensity difference like in the experiment, where the Bragg peak is spread over a finite wavevector range. In addition, the low wavevector broadening of the peak takes an order of magnitude longer to recover in the experiment because it cannot exceed the finite penetration depth of the pump.

Figure \ref{fig:simulations}(b) shows a series of isosurfaces of the order parameter, showing the spatiotemporal evolution of the vortex strings where the order parameter is topologically locked to zero. Upon quenching, the vortex strings are confined to the domain wall. As time evolves from the quench, the vortex strings diffuse throughout the plane and annihilate, with most strings annihilated by $t = 53$ ps. Figure \ref{fig:simulations} (c) shows the time-dependent structure factors $S(\bm{q},t)$ collapse onto a single curve when scaling with $\gamma = \frac{1}{2}$ critical exponent. The discrepancy between this critical exponent and our experimental observation of $\gamma \approx \frac{1}{3}$ is a known effect in Ginzburg-Landau simulations \cite{mondello_scaling_1992,orenstein_dynamical_2025}. Figure \ref{fig:simulations}(c) shows the scattering intensity in the out of plane direction fits to a $\eta = -4$ power-law, with $\eta = -2$ dependence for high $q$ due to the persistent phase modes (coming from the part of region I in fig. \ref{fig:simulations} (a)  that overlaps with region II). This is a direct result of the initial spatial orientation of the vortex strings to the plane of the domain wall.

Figure \ref{fig:simulations} (d) shows the real part of the order parameter along the $xz$-plane corresponding to the formation and annihilation of topological defects at the domain wall. Figure \ref{fig:simulations} (e) shows the evolution of the real part of the order parameter $\phi$ along the $z$-dimension, showing that the coarsening along the $z$-axis continues across multiple time and length scales, even after the vortex strings are mostly annihilated.


Three key features of the data appear in both the simulation and experiment, and illustrate important parallels and differences between ultrafast photoexcitation and a uniform quench \cite{kibble_topology_1976,zurek_cosmological_1985,dolgirev_self-similar_2020}. First, the universal scaling and critical exponents are capable of predicting dynamics of the system at times much longer than the quench timescale, e.g. longer than one picosecond. As shown in fig. \ref{fig:data}(a), the power-law behavior of the domain coarsening appears at timescales longer than 10 ps. At very short timescales, the ballistic motion of the amplitude and phase modes of the CDW are significant, leading to distinct control mechanisms at these timescales. 

Second, the power-law scaling of the intensity as a function of wavevector is different in a uniform quench ($\eta = -5$) and from the previously observed scaling in LaTe$_3$ \cite{orenstein_dynamical_2025} than in our simulations and experiment ($\eta = -4$). We interpret this difference as a reduction in the effective dimensionality of the vortices, imposed by the finite penetration depth of the pump. This finite penetration depth produces a single, 2D domain wall at short times. This domain wall subsequently decays into topological defects which eventually annihilate completely (fig. \ref{fig:simulations}(b)), but the initial \emph{orientation} of the CDW phase produces an effectively reduced dimension of the problem --- all the vortex strings are preferentially oriented in a single plane.

Third, the finite penetration depth produces a defined length scale which interacts with the variable length scale of the domain coarsening in ways that are not apparent in a uniform, infinite system. This is shown in fig. \ref{fig:data}(a) and fig. \ref{fig:simulations}(a), from $t = 1$ ns to $t = 1$ $\mu$s in experiment, when the finite wavevector domain-wall scattering appears to disappear into the Bragg peak.  In real space, this corresponds to the length scale of vortex strings approaching the distance between the domain wall and the crystal surface. Critically, there is still a time-resolved signal apparent in both experiment and simulation, indicating that the system has not yet returned to an equilibrium configuration, which occurs more than an order of magnitude in time later. 

These observations show fundamental mechanistic differences between the coarsening dynamics of photoinduced phase transitions compared to a uniform quench. For photoinduced phase transitions in CDWs, \emph{neither} a fully coherent framework \cite{trigo_coherent_2019,maklar_nonequilibrium_2021} nor a uniform quench framework \cite{orenstein_dynamical_2025,lal_universal_2020,mondello_scaling_1990,danz_ultrafast_2021,dolgirev_self-similar_2020,zong_role_2021,kogar_light-induced_2020} describe key features of the phase transition. The finite timescale of the quench, the finite length scale of the penetration depth, and the finite timescale of the amplitude mode must all be accounted for in order to accurately predict the pathway to the final state of the system. 

In conclusion, we demonstrate that using ultrafast x-rays to resolve the finite-wavevector scattering after strong photoexcitation of a CDW system reveals the complex interplay of finite penetration depth, finite quench time, and topological vortex string coarsening during CDW relaxation. This interplay produces topologically distinct protected intermediate structures at timescales orders of magnitude longer than the electronic recovery time. These structures could offer novel routes to stabilizing light-induced states of matter. More layers of control over material dynamics could be achieved with further spatial structuring of light, by e.g. transient gratings \cite{li_nanoscale_2025,rouxel_hard_2021, bencivenga_four-wave_2015} or plasmonic nanostructures \cite{wang_plasmonic_2007}.

\begin{acknowledgments}
G.O., R.A.D, V.K., and M.T. were supported by the US Department of Energy, Office of Science, Office of Basic Energy Sciences through the Division of Materials Sciences and Engineering under Contract No. DE-AC02-76SF00515. We acknowledge the Paul Scherrer Institute, Villigen, Switzerland for provision of free-electron laser beamtime at the Bernina instrument of the SwissFEL ARAMIS branch. 
\end{acknowledgments}

\bibliography{ErTe3_references}

\end{document}